\documentclass[sigconf]{acmart}
\acmConference[EASE 2026]{The 30th International Conference on Evaluation and Assessment in Software Engineering}{9–12 June, 2026}{Glasgow, Scotland, United Kingdom}
\usepackage{graphicx}
\usepackage{amsmath}
\usepackage{tabularx}   
\usepackage{multirow}
\usepackage{forest}
\usepackage{subcaption}   
\usepackage{caption}
\usepackage{pgfplots}
\usepackage{adjustbox}
\usepackage{comment}
\usepackage[T1]{fontenc}
\pgfplotsset{compat=1.18}
\newcolumntype{L}{>{\raggedright\arraybackslash}X}
\newcolumntype{Y}{>{\centering\arraybackslash}X} % 非数值列自动拉伸

\usepackage{forest}
\usepackage{subcaption}
\usepackage{tikz}
\usetikzlibrary{arrows.meta,calc}
\forestset{
  unifiedAST/.style={
    for tree={
      draw,
      rounded corners=2pt,
      align=center,
      font=\ttfamily\scriptsize,
      minimum height=4.2mm,
      inner sep=1.2pt,
      l sep=8pt,        % 层距（父子）
      s sep=6pt,        % 兄弟间距
      edge={-Latex, line width=0.5pt, draw=black!65},
      edge path'/.style={
        !u.parent anchor=children
        ..controls +(.7em,0) and +(-.7em,0)..
        (.child anchor)
      },
    },
  },
  nodetype/.style={fill=blue!6, draw=blue!35},
  token/.style={fill=orange!8, draw=orange!45},
}

\AtBeginDocument{%
  }

\setcopyright{acmlicensed}
\copyrightyear{2018}
\acmYear{2018}
\acmDOI{XXXXXXX.XXXXXXX}
\acmISBN{978-1-4503-XXXX-X/2018/06}

\begin{document}

\title{Bridging the Programming Language Gap: Constructing a Multilingual Shared Semantic Space through AST Unification and Graph Matching}

\author{Junhao Chen}
\email{chenjjhh@nuaa.edu.cn}
\affiliation{%
  \institution{Nanjing University of Aeronautics and Astronautics}
  \country{China}
}

\author{Jingxuan Zhang}
\authornote{Corresponding author.} 
\email[Corresponding author]{jxzhang@nuaa.edu.cn}
\affiliation{%
  \institution{Nanjing University of Aeronautics and Astronautics}
  \country{China}
}

\author{Jian He}
\email{hejian7175@sina.com}
\affiliation{%
  \institution{Shanghai Aerospace Electronic Technology Institute}
  \country{China}
}

\author{Yixuan Tang}
\email{tangyixuan@nuaa.edu.cn}
\affiliation{%
  \institution{Nanjing University of Aeronautics and Astronautics}
  \country{China}
}

\author{Weiqin Zou}
\email{weiqin@nuaa.edu.cn}
\affiliation{%
  \institution{Nanjing University of Aeronautics and Astronautics}
  \country{China}
}

\renewcommand{\shortauthors}{Trovato et al.}

\begin{abstract}
The lexical and syntactic disparities among different programming languages (e.g., Java and Python) pose significant challenges for multi‐language software engineering tasks such as cross‐language code clone detection and code retrieval, since queries or code snippets written in one programming language often fail to match equivalent artifacts in another. To bridge this gap between different programming languages, we proposed a novel approach to construct a multi-language shared semantic space, in which functionally equivalent source code written in different programming languages are close to each other. In this approach, we first map the Abstract Syntax Tree (AST) node labels of the code snippets written in different programming languages into a unified label set, thus compressing high-dimensional language-specific tokens into a common embedding space. Then, we employ a Graph Matching Network (GMN) to encode the paired AST graphs into ``semantic vectors'' that capture functional equivalence between programming languages in a unified code vector space. In such a way, we can eliminate the differences in syntax between different programming languages. To validate the effectiveness of this approach, we apply it to two downstream tasks, including cross‐language clone detection and cross‐language code retrieval. Experiments demonstrate that our approach substantially outperforms the state-of-the-art baselines in cross-language clone detection, improving Precision from 95.62\% to 99.94\%, Recall from 97.72\% to 99.92\%, and F1 score from 96.94\% to 99.93\%. In terms of cross-language code retrieval, our approach raises the average Mean Reciprocal Rank (MRR) from 0.4909 to 0.5547, showing an absolute gain of 0.0638 ($\approx 13\%$ relative improvement), which demonstrates its superior ability to rank correct code snippets high across multiple programming languages.
\end{abstract}




%%\received{20 February 2007}
%%\received[revised]{12 March 2009}
%%\received[accepted]{5 June 2009}

\maketitle

\section{Introduction}

In the modern software engineering practice, polyglot development has become commonplace. Large-scale software projects often incorporate multiple programming languages. For example, a software project may employ JavaScript on the front end, Java or Python on the back end, and C++ or C\# for system components \cite{mussbacher2024polyglot}. Although this multi-language strategy employed in software projects can fully leverage the strengths of each programming language, it also poses significant challenges for code comprehension and maintenance \cite{yang2024multi}. In particular, for cross-language code understanding tasks, such as cross-language code clone detection and code retrieval, developers frequently need to find functionally equivalent or semantically similar code snippets across different programming languages to support code refactoring, code reuse, and bug localization \cite{yu2019neural, mehrotra2023improving, wang2022unified}. However, because programming languages differ markedly in their syntax in both the shapes of Abstract Syntax Trees (ASTs) and in the coding styles, directly comparing or retrieving the source code across different programming languages often fails to produce satisfactory results \cite{perez2019cross}.

Cross-language code understanding and manipulation face two main challenges. First, different programming languages have distinct Abstract Syntax Tree (AST) node labels and structures, making direct comparison of ASTs difficult. Even with identical functionality, ASTs can vary significantly due to syntactic differences, library calls, and language-specific idioms. Although some methods aim to unify ASTs by mapping heterogeneous node types, the second challenge persists: functionally equivalent ASTs may still differ semantically due to subtle implementation details. Therefore, it is crucial to model cross-graph semantic alignment at the node level to bring semantically equivalent code closer in a shared semantic space.

To address these challenges, we propose a two-phase approach for constructing a cross-language shared semantic space \cite{li2019graph, wang2020detecting, zhang2019novel}. In the first phase, we use unified AST abstraction to eliminate language-specific differences, mapping ASTs to a common set of "universal semantic labels" \cite{lei2022deep}. This phase prepares ASTs for the subsequent processing. However, even after unification, functionally equivalent ASTs may diverge in semantics. To overcome this, the second phase employs a Graph Matching Network (GMN) to capture "graph-to-graph interactions" at the node level, enabling semantic alignment across graphs and pulling similar ASTs closer in a shared space.

Using the shared semantic space generated by the unified AST abstraction and GMN-based alignment, we apply our approach to two tasks, i.e., cross-language clone detection and code retrieval. For clone detection, we use a contrastive learning framework to train a classifier that distinguishes true code clones from similar non-clones. For code retrieval, we index embeddings of large-scale code snippets and use cosine similarity to retrieve semantically equivalent snippets across languages.

We evaluate our approach on the publicly available Google Code Jam \cite{codejam}, AtCoder \cite{atcoder}, and XLCoST \cite{zhu2022xlcost} benchmark datasets, covering multiple programming languages, including Java, C++, Python, and C\#. In the cross-language clone detection, our approach achieves Precision of $99.94\%$, Recall of $99.92\%$, and F1 score of $99.93\%$, surpassing the state-of-the-art baselines such as FSD-CLCD ($95.62\%/97.72\%/96.94\%$). In the cross-language code retrieval, our approach improves the average Mean Reciprocal Rank (MRR) from $0.4909$ to $0.5547$, and achieves Precision@4 above $91\%$ in most code retrieval scenarios with different source and target programming languages. These results demonstrate the effectiveness of our approach in supporting both the classification and retrieval tasks across diverse programming languages.

In summary, this paper makes the following three contributions.
\begin{itemize}
    \item We propose a cross-language unified AST abstraction that maps AST node labels of different languages to a common set of universal labels, reducing syntax gaps.
    \item We introduce a GMN-based cross-graph semantic alignment model, using cross-graph attention to align semantically equivalent code snippets in a shared semantic space.
    \item We open the replication package to the public for future extensions\footnote{\url{https://github.com/CenOspreay/BPLangGap}}. Additionally, we apply our approach to cross-language code clone detection and retrieval, achieving superior results compared to state-of-the-art methods.
\end{itemize}

%The remainder of this paper is organized as follows. Section 2 presents the motivation. Section 3 introduces the framework of our approach. Section 4 presents the experiments on the cross-language clone detection and code retrieval. Section 5 discusses the visualization results and Section 6 reviews the related work. Finally, Section 7 concludes this paper.

\section{Motivation}

In the cross-language code understanding and manipulation, there are two main challenges, including inconsistent AST node labels as well as hierarchies of the code snippets across different programming languages and vastly different AST shapes exhibited in functionally identical code snippets, making deep semantic correspondence of code snippets from different programming languages hard to be captured. We provide several examples to illustrate what these challenges are and how we can tackle them.

\subsection{Unified AST Abstraction}

ASTs are widely used as intermediate representations in cross-language code understanding and manipulation. However, AST node labels and structural patterns are inherently language-specific. Even for functionally equivalent programs, their AST shapes and node-type vocabularies can differ substantially across languages \cite{wang2022unified, swilam2023cross, perez2019cross, song2024revisiting}. Consequently, AST features designed for a single language are difficult to generalize to multilingual settings.

\begin{figure}[!t]
\centering
\resizebox{0.8\linewidth}{!}{
\begin{subfigure}[t]{0.49\linewidth}
\centering
\hspace*{-2.0em}
\begin{forest}
unifiedAST
[VariableDeclaration\\{(sum, list)}, nodetype
  [ForStatement, nodetype
    [Initialization\\{(i=0)}, token]
    [Condition\\{(i<list.size())}, token]
    [Update\\{(i++)}, token]
    [Body, nodetype
      [Assignment\\{(sum = sum + get(i))}, token]
    ]
  ]
]
\end{forest}
\caption{Java code and its AST}
\end{subfigure}
\hfill
\begin{subfigure}[t]{0.49\linewidth}
\centering
\hspace*{2.0em}
\begin{forest}
unifiedAST
[Module\\{(sum, lst)}, nodetype
  [For, nodetype
    [Target\\{(elem)}, token]
    [Iter\\{(lst)}, token]
    [Body, nodetype
      [AugAssign\\{(sum += elem)}, token]
    ]
  ]
]
\end{forest}
\caption{Python code and its AST}
\end{subfigure}

\vspace{-0.35em}
}
\caption{Structural comparison of AST nodes}
\label{fig:java-python}
\vspace{-0.5em}
\end{figure}
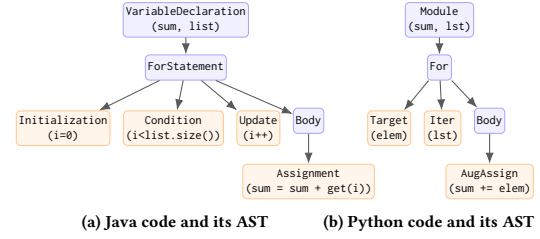

Figure~\ref{fig:java-python} illustrates this challenge. Both code snippets implement the same task, i.e., iterating over a list and accumulating the sum. However, their corresponding ASTs are organized differently due to distinct syntax and idioms. The Java code uses an index-based \texttt{ForStatement} with explicit initialization/condition/update and an \texttt{Assignment}, whereas the Python code adopts an iterator-style \texttt{For} loop and represents accumulation via \texttt{AugAssign}. Such label and topology mismatches make direct cross-language alignment unreliable, as semantically equivalent fragments may appear unrelated under naive node-label or subtree-structure matching \cite{mehrotra2023improving, song2024revisiting}.

To address this issue, we design a unified AST abstraction process that projects language-specific ASTs into a shared node label space. By mapping heterogeneous node-type vocabularies to a compact set of universal labels and normalizing key structural variations, our abstraction process reduces syntactic noise while preserving fine-grained functional cues and the essential AST topology. This enables AST fragments from different languages to be embedded into the same feature space and compared directly for the cross-language alignment.

\usetikzlibrary{arrows.meta}

% 定义统一的AST样式
\forestset{
  unifiedAST/.style={
    for tree={
      draw,
      rounded corners=2pt,
      align=center,
      font=\ttfamily\scriptsize,
      minimum height=4.2mm,
      inner sep=1.2pt,
      l sep=8pt,        % 层距（父子）
      s sep=6pt,        % 兄弟间距
      edge={-Latex, line width=0.5pt, draw=black!65},
      % 让边是“先水平后竖直”的正交连线，更清晰紧凑
      edge path'/.style={
        !u.parent anchor=children
        ..controls +(.7em,0) and +(-.7em,0)..
        (.child anchor)
      },
    },
  },
  nodetype/.style={fill=blue!6, draw=blue!35},
  token/.style={fill=orange!8, draw=orange!45},
}

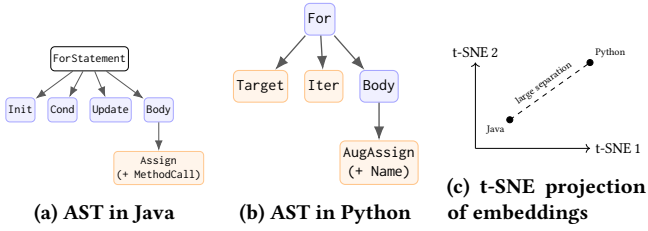
\begin{figure}[t]
\centering

\begin{subfigure}[b]{0.31\linewidth}
\centering
\begin{adjustbox}{max width=\linewidth}
\begin{forest}
unifiedAST
[ForStatement
  [Init, nodetype]
  [Cond, nodetype]
  [Update, nodetype]
  [Body, nodetype
    [Assign\\(+ MethodCall), token]
  ]
]
\end{forest}
\end{adjustbox}
\caption{AST in Java}
\end{subfigure}
\hfill
\begin{subfigure}[b]{0.31\linewidth}
\centering
\begin{adjustbox}{max width=\linewidth}
\begin{forest}
unifiedAST
[For, nodetype
  [Target, token]
  [Iter, token]
  [Body, nodetype
    [AugAssign\\(+ Name), token]
  ]
]
\end{forest}
\end{adjustbox}
\caption{AST in Python}
\end{subfigure}
\hfill
\begin{subfigure}[b]{0.31\linewidth}
\centering
\begin{adjustbox}{max width=\linewidth}
\begin{tikzpicture}[scale=0.60]
  \draw[->] (0,0) -- (4,0) node[right] {t-SNE\,1};
  \draw[->] (0,0) -- (0,3) node[above] {t-SNE\,2};
  \filldraw (1.2,1.0) circle (3pt) node[below left] {\scriptsize Java};
  \filldraw (4.0,3.0) circle (3pt) node[above right] {\scriptsize Python};
  \draw[dashed, semithick] (1.2,1.0) -- (4.0,3.0)
    node[midway, above, sloped] {\scriptsize large separation};
\end{tikzpicture}
\end{adjustbox}
\caption{t-SNE projection of embeddings}
\end{subfigure}

\caption{Java and Python ASTs that are structurally similar but mapped far apart in the semantic space}
\label{fig:ast_semantic_drift}
\end{figure}

\subsection{Cross-Graph Semantic Alignment}

Although the unified AST abstraction removes most discrepancies at the node level of ASTs, representations of code snippets from different languages can still drift apart in semantic space (see Figure ~\ref{fig:ast_semantic_drift}). For example, even if two ASTs implement the same functionality such as iterating over a list and summing its elements, their encoded vectors may lie far apart after projection into a two-dimensional semantic space, illustrating the phenomenon of semantic drift. To further bring functionally equivalent cross-language ASTs close to each other in the semantic space, we need a method that explicitly compares and aligns two graph structures of ASTs.

\begin{figure*}[t]
  \centering
  \includegraphics[width=0.95\textwidth]{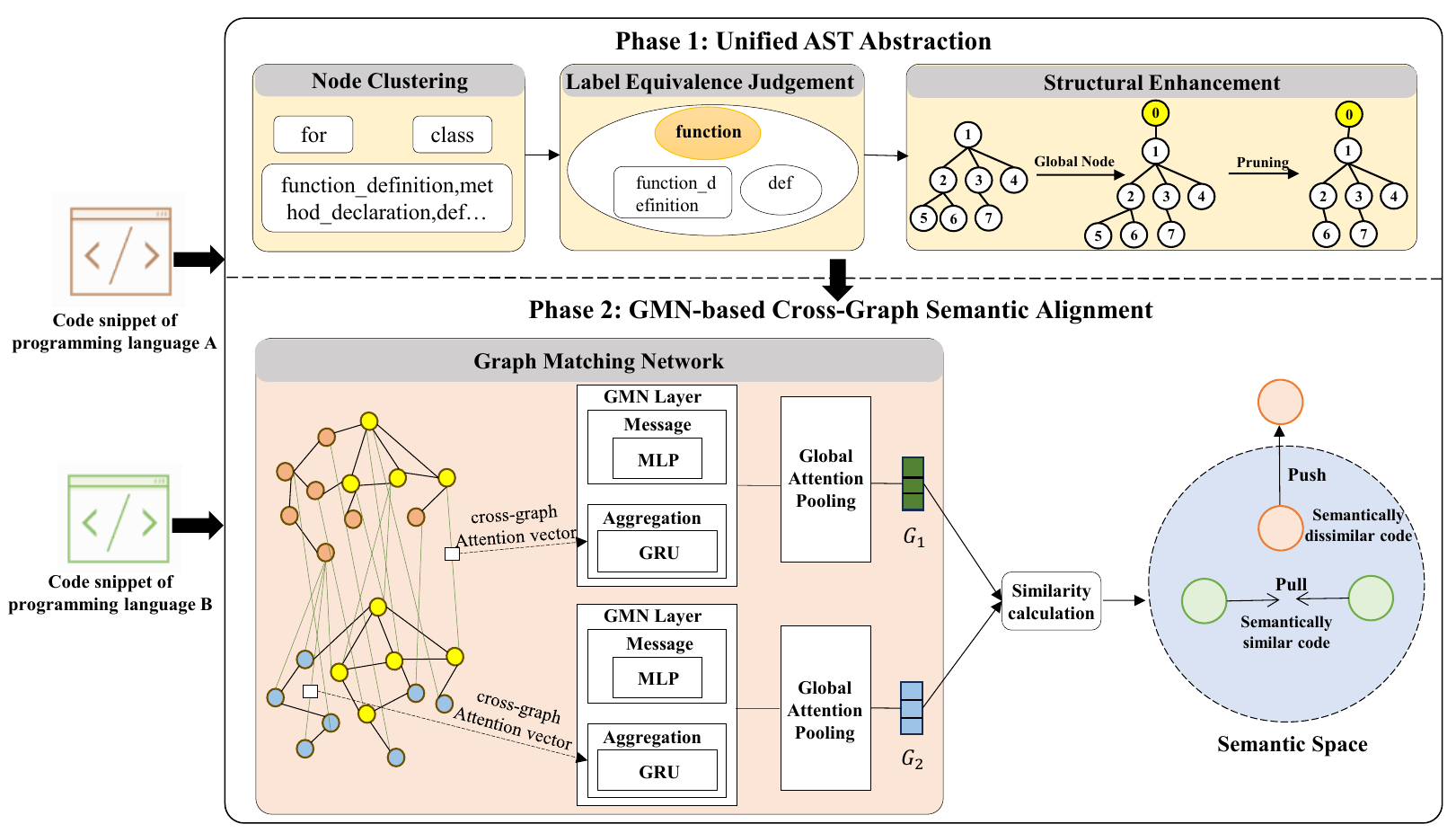}
  \caption{Overall framework of our approach}
  \label{fig:framework}
\end{figure*}

Even though there are several Graph Neural Networks (GNNs) in the literature, they can only aggregate information within one graph and cannot directly model interactions between a pair of graphs \cite{scarselli2008graph, wu2020comprehensive}. Recently, the Graph Matching Network (GMN) is proposed, which offers exactly the several capabilities, including cross-graph attention interactions, intra-graph iterative updates, and global representation aggregation. In the cross-graph attention interactions, each node in AST$_1$ attends to every node in AST$_2$ (and vice versa) during encoding, enabling information flow between the two graphs \cite{li2019graph}. In terms of the intra-graph iterative updates, after aggregating cross-graph messages, GMN uses a graph-level GRU to iteratively update the embedding of each node, so it retains its local structural context while incorporating semantic cues from the other graph \cite{cheng2022cross}. In addition, in the global representation aggregation, following several rounds of interaction, GMN applies the global attention pooling (or another readout) on each graph to produce a single vector per graph, thus aware of its counterpart, making it well suited for similarity measurement or classification \cite{chen2021learning}.

By combining the unified AST abstraction with the GMN-based cross-graph semantic alignment, we can pull together functionally equivalent code snippets from different programming languages in a shared semantic space, significantly improving the performance of cross-language code understanding and manipulation.

\section{Framework}

Based on the motivation, we propose a novel approach consisting of two phases, i.e., unified AST abstraction and GMN-based cross-graph semantic alignment, as illustrated in Figure~\ref{fig:framework}. In the first phase, our approach processes the code snippets from different languages (e.g., Java, C++, and Python) into unified ASTs. Specifically, all AST nodes are first \textbf{classified and mapped} into a unified set of labels to ensure consistent semantic representation (node classification). Then, semantically equivalent labels such as \texttt{FunctionDeclaration} in Java and \texttt{FunctionDecl} in C++ are merged through \textbf{label equivalence judgment}, ensuring the cross-language uniformity. To further normalize the structure, \textbf{structural enhancement} is applied by pruning non-essential nodes (e.g., comments and empty statements) and inserting a \textbf{global root node} as a shared anchor for subsequent alignment \cite{wang2022unified, sun2023abstract}. After these processing, each code snippet is represented as a structurally enhanced AST over the unified label set.

The second phase performs GMN-based semantic alignment. It takes two unified ASTs as input and, through four steps, i.e., node feature encoding (MLP), intra-graph aggregation updates (GRU), cross-graph attention interaction (cross-graph attention vector), and graph-level representation generation (Global Attention Pooling), to project both ASTs into the same vector space, yielding a pair of “mutually aware” semantic vectors \cite{cheng2022cross,li2019graph}. The cosine similarity score between these vectors then quantifies the functional similarity of the two code snippets.

\subsection{Unified AST Abstraction}

In this phase, we transform the raw ASTs of code snippets from different programming languages through a sequence of processing steps so that they all map onto a single set of universal semantic labels while preserving essential attributes and structural information.

\subsubsection{Node Clustering}

Our goal is to build a universal node label set for ASTs that assigns semantically equivalent node labels from different programming languages to the same labels. To avoid the incompleteness of manually curated mappings, we construct the unified label space in an automated and grammar-driven manner.

For each language $L$, we obtain a complete inventory of \emph{named} AST node labels by exporting the parser/grammar metadata (i.e., the node specification derived from the programming language grammar). This yields a language-specific node set $\mathcal{T}_L$, and the multilingual universe is defined as $\mathcal{T}=\bigcup_L \mathcal{T}_L$, ensuring comprehensive coverage including rare node labels that may not appear in the training data.

Next, for each node label $t\in\mathcal{T}_L$, we build a language-agnostic signature by combining (i) normalized type-name features and (ii) grammar-structural schema features (field names, admissible child types, optional/repeatable constraints, and typical arity). When we align AST node labels across different programming languages, we find that some nodes, such as `for' and `while' loops, may have different syntactic representations but share similar semantics. Hence, we enhance our approach by integrating CodeBERT to assist in identifying equivalent nodes, despite differences in their syntactic structure. The procedure uses CodeBERT to encode the features of node labels from different languages into vector representations, i.e., embeddings. The embeddings are then compared and calculated using the cosine similarity. If the similarity between the embeddings of two different node labels exceeds a threshold of 0.75, they are considered equivalent and mapped to the same universal label. We conducted preliminary experiments to validate this threshold and found that a cosine similarity of 0.75 achieved the best balance between precision and recall in capturing equivalent node labels. After we calculate the similarities between different node labels of different programming languages, we apply connected-components clustering (Union-Find) to group labels into universal clusters. As a result, node labels with high similarities are clustered into the same universal category (e.g., calls, loops, assignments, and conditionals), resulting in a compact universal label set shared across languages.

%This semantic alignment approach allows our approach to make more nuanced decisions beyond simple syntactic matching. For example, the model can identify that a 'for' loop in one language and a 'while' loop in another language, despite having different syntactic structures, are often used to accomplish the same task and can thus be classified under the same universal node label.

\subsubsection{Label Equivalence Judgment}

After the initial alignment, we further consolidate the universal labels automatically to reduce redundancy caused by parser granularity and naming conventions. Specifically, we merge clusters that are structurally compatible under the grammar schemas and exhibit high signature similarity. To reduce noise, we detect infrequent or non-essential node labels using frequency and context-heterogeneity criteria, and map them to an \texttt{Other} category. This yields a compact yet robust universal label set without manual intervention. The integration of CodeBERT ensures that only semantically consistent and relevant node labels are grouped together, improving the accuracy of our label set while avoiding misalignments that could arise from superficial syntactic differences.

\subsubsection{Structural Enhancement}

We enhance the overall AST structure to make them clean and semantically focused. First, we insert a global root node at the top of each AST. This artificial root does not correspond to any concrete language construct but serves to unify independent control‐flow or declaration subtrees under a single parent \cite{zhang2024fsd, hovemeyer2016control}. Specifically, all top‐level constructs, such as function declarations and module entry points, become the direct children of this global root node. This guarantees that, regardless of their original order or nesting in different programming languages, these high‐level nodes are collected under a common anchor, which greatly facilitates whole‐graph alignment \cite{wang2024segmn}. Next, we protect key nodes by preserving their full structure and attributes. The key nodes \cite{wang2024segmn} include:

\begin{enumerate}
    \item Control flow nodes (loops like FOR\_LOOP and conditional branches like IF\_COND), which drive the high level logic of the code snippet.
    \item Declaration nodes (e.g., FUNC\_DEF, VAR\_DECL), which convey the abstract structure and type information of the code snippet.
    \item Critical attribute nodes (e.g., constants such as CONST\_NUM or CONST\_STR, and type references like TYPE\_REF), which carry the runtime semantics or type constraints of the code snippet.
\end{enumerate}

The other nodes, such as auxiliary syntactic sugar, modifiers, intermediate wrappers, or language-specific formatting constructs, are regarded as unimportant nodes. These nodes are typically parser-dependent artifacts that do not alter the functionality of the code snippet but enlarge the AST and introduce \emph{linguistic feature noise}, as also discussed in the prior work \cite{zhang2024fsd}. If they are retained, they increase the cross-language divergence since different programming languages implement the same functionality with very different amounts of syntactic decoration. Pruning them does not break the completeness of the underlying code semantics, since the protected key nodes already preserve control flow, declarations, and constant values, which are sufficient to reconstruct the functionality of the code snippet.

Following the pruning strategy of FSD-CLCD \cite{zhang2024fsd}, the process is not a one-time removal of all non-essential nodes. Instead, pruning is performed under the following key node protection rules:

\begin{itemize}
    \item All edges connected to global nodes are preserved. 
    \item All edges of key nodes and their first-order neighbors are preserved. 
    \item The remaining edges are subject to random deletion at a fixed pruning ratio (0.4 in this study). 
\end{itemize}

This random but ratio-controlled deletion avoids excessive structure loss while reducing the node-size disparity across programming languages. Empirical results confirm that these rules eliminate language-specific noise, balance graph sizes, and improve the robustness of cross-language alignment without harming the semantic integrity of the code snippet \cite{zhang2024fsd}.

\subsection{GMN Semantic Space}

After obtaining two ASTs of the code snippets from different programming languages that have been unified and retained necessary attributes and structure, we project them into a common semantic space to compute the cross-language similarity. To this end, we employ a GMN \cite{li2019graph} to encode and align both ASTs simultaneously.
    
\subsubsection{Node Feature Encoding}

For each node in each AST, we start with an initial vector formed by concatenating its universal type embedding and its attribute embedding. Formally, let $y_i$ denote the unified node label of node $i$, and $(w_{i,1},\dots,w_{i,m_i})$ the tokenized sequence of its attributes (e.g., identifier names or constant values). The universal type embedding is obtained from a trainable lookup table $\mathbf{E}_{\mathrm{type}} \in \mathbb{R}^{|\mathcal{L}|\times d_t}$:
\begin{equation}
    \mathbf{t}_i = \mathbf{E}_{\mathrm{type}}[y_i] \in \mathbb{R}^{d_t}.
\end{equation}

The attribute embedding is computed by averaging the embeddings of its tokens using $\mathbf{E}_{\mathrm{attr}} \in \mathbb{R}^{|\mathcal{V}|\times d_a}$:
\begin{equation}
    \mathbf{v}_i = \frac{1}{m_i}\sum_{j=1}^{m_i}\mathbf{E}_{\mathrm{attr}}[w_{i,j}] \in \mathbb{R}^{d_a}.
\end{equation}

Finally, the initial node vector is defined as the concatenation of the two parts:
\begin{equation}
    \mathbf{h}_i^{(0)} = [\,\mathbf{t}_i \,;\, \mathbf{v}_i\,] \in \mathbb{R}^{d_t+d_a}.
\end{equation}

In the first step of the GMN, these vectors are passed through a Multi-Layer Perceptron (MLP) and a two-layer fully connected network with the ReLU activations, which project the node features into a unified dimensional space. We then apply LayerNorm and Dropout to produce the initial node representations ${{\mathbf{z}}_i^{{(0)}}\mathrm{}}$ \cite{duong2019node, maurya2022simplifying}.
\begin{align}
z_i^{(0)} = \mathrm{LayerNorm}\bigl(\mathrm{ReLU}\bigl(W_2\bigl(\mathrm{ReLU}(W_1 h_i^{(0)} + b_1)\bigr) + b_2\bigr)\bigr)
\end{align}

This process integrates the heterogeneous attribute information, ensuring that all nodes share a consistent vector‐dimensional format \cite{zhao2024ra}.

\subsubsection{Cross-Graph Attention Interaction}

Leveraging a cross-graph attention mechanism, the GMN enables nodes from both ASTs to “see” each other during the encoding. Specifically, for each node \(i\) in AST$_1$, we calculate the attention scores against each node \(j\) in AST$_2$.
\begin{align}
e_{ij} = \mathrm{LeakyReLU}\!\bigl(\mathbf{a}^\top [\,\mathbf{z}_i^{(t)} \,\|\, \mathbf{z}_j^{(t)}]\bigr)
\end{align}

Then, we normalize over the \(j\) dimension to obtain the attention weights $\alpha_{ij}$:
\begin{align}
\alpha_{ij} = \frac{\exp\bigl(e_{ij}\bigr)}{\sum_{j'} \exp\bigl(e_{ij'}\bigr)}
\end{align}

Thus, by the weighted summation, we aggregate the features of all nodes in AST$_2$ to generate the cross‐graph context $\tilde{c}_i$.
\begin{align}
\tilde{c}_i = \sum_{j \in \mathrm{Nodes}(\mathrm{AST}_2)} \alpha_{ij} \, z_j^{(t)}
\end{align}

For each node \(j\) in AST\(_2\), we perform the same operation over all nodes in AST\(_1\) to obtain its cross‐graph context $\tilde{c}_j$. This interactive attention ensures that every node in AST\(_1\) receives the global information from AST\(_2\), and vice versa, thereby uncovering deep potential correspondences between the two ASTs \cite{li2019graph, ling2021multilevel}.

\subsubsection{Intra-graph Aggregation and Update}

In each iteration, the GMN uses a Gated Recurrent Unit (GRU) to fuse and update the cross-graph context with the local graph information. For node \(i\) in AST\(_1\), we concatenate its current feature $\mathbf{z}_i^{(t)}$ and the cross-graph context $\tilde{\mathbf{c}}_i$:
\begin{align}
\mathbf{u}_i = \bigl[\mathbf{z}_i^{(t)} \,\|\, \tilde{\mathbf{c}}_i\bigr] \in \mathbb{R}^{2d}
\end{align}

Then, based on the concatenated vector \(\mathbf{u}_i\), we aggregate the features of its neighbors according to the adjacency of the graph:
\begin{align}
\bar{n}_i = \frac{1}{\lvert \mathcal{N}(i)\rvert} \sum_{k \in \mathcal{N}(i)} \mathbf{z}_k^{(t)}
\end{align}

The GRU, employing its update‐gate and reset‐gate mechanisms, takes both the concatenated vector \(\mathbf{u}_i\) and the neighbor‐aggregated vector \(\bar{\mathbf{n}}_i\) as the input. It outputs the updated node representation as follows:
\begin{align}
\mathbf{z}_i^{(t+1)} = \mathrm{GRU}\bigl(\mathbf{u}_i, \bar{\mathbf{n}}_i\bigr)
\end{align}

Thus, through multiple rounds of cross-graph attention and intra-graph GRU updates, the model gradually learns a comprehensive embedding for each node in ASTs that captures both its own syntactic structure and semantic cues from the other graph. After \(T\) iterations, the representations of every node in AST$_1$ and AST$_2$ have fully fused cross-graph information \cite{li2019graph}.

\subsubsection{Graph-level representation generation}

To obtain global vector representations for both ASTs, the GMN employs the global attention pooling. Specifically, for AST$_1$, it computes a weighted sum of the final representations of all the nodes according to the trainable pooling weights as follows:
\begin{align}
\gamma_i \;=\; \frac{\exp\bigl(\mathbf{w}^\top \mathbf{z}_i^{(T)}\bigr)}
{\sum_{k \in \mathrm{Nodes}(\mathrm{AST}_1)} \exp\bigl(\mathbf{w}^\top \mathbf{z}_k^{(T)}\bigr)}, 
\mathbf{v}_1 \;=\; \sum_{i \in \mathrm{Nodes}(\mathrm{AST}_1)} \gamma_i \,\mathbf{z}_i^{(T)}
\end{align}

The nodes in AST$_2$ undergo the same pooling procedure to produce attention weights and a global representation vector. Since the embedding of each node has fully interacted with the other graph during the iterations, these two vectors intrinsically encode deep, mutually aware semantic information and can be directly used for the cosine similarity computation as follows.
\begin{align}
\mathrm{sim}(\mathbf{v}_1, \mathbf{v}_2) \;=\; \frac{\mathbf{v}_1^\top \mathbf{v}_2}{\|\mathbf{v}_1\| \,\|\mathbf{v}_2\|}
\end{align}

We evaluate our approach on cross-language code clone detection and retrieval. For clone detection, we compare against state-of-the-art approaches to show the benefit of our shared semantic space. For retrieval, we train with negative sampling and a temperature-scaled cross-entropy loss. Overall, the results demonstrate strong effectiveness and generalization.

\section{Applications of Our Approach}

This section evaluates our approach on two representative cross-language program understanding tasks, namely cross-language code clone detection and cross-language code retrieval. To guide our empirical study and make the evaluation objectives explicit, we formulate the following research questions:

\begin{itemize}
	\item \textbf{RQ1:} Can our approach achieve better results than the baseline approaches?
	\item \textbf{RQ2:} Is the hard-negative instance construction strategy effective in our approach?
	\item \textbf{RQ3:} What is the effect of each module on the performance of our approach?
\end{itemize}

To answer \textbf{RQ1}, we compare our approach with representative baselines under both multilingual and one-to-one cross-language settings for clone detection, and report evaluation metrics (e.g., MRR and Precision@K) for code retrieval. To answer \textbf{RQ2}, we conduct controlled experiments by enabling/disabling the hard-negative strategy while keeping other settings unchanged. To answer \textbf{RQ3}, we perform an ablation study by systematically removing or replacing key modules and analyzing the performance changes.

\subsection{The Benchmark Dataset}

\textbf{Code clone detection.} We use the same benchmark dataset for cross-language code clone detection published by Tao et al. \cite{tao2022c4}, which is collected from two public programming contest platforms, i.e., Google Code Jam and AtCoder. In the code clone detection task, code snippets to the same programming problem in different programming languages are treated as code clone pairs (positive pairs), and code snippets targeted towards different programming tasks as non-code clone pairs (negative pairs). The dataset comprises 88,568 code snippets covering 1,380 distinct programming tasks, with an average of 64 code snippets per programming task. We split the dataset based on the programming tasks into training, validation, and test sets in an 8:1:1 ratio, namely 1,100 programming tasks for training and 140 each for validation and testing, ensuring no overlap of the programming tasks across splits.

%We construct the training, validation, and test sets with a 1:1 ratio of positive to negative samples under a contrastive-learning framework \cite{hadsell2006dimensionality}. For the negative code clone pair construction, we employ the hard-negative construction strategy, i.e., the most non-matching code snippet (with the largest distance) is selected and combined with a specific code snippet to form the negative code clone pair. This strategy maintains the 1:1 positive-to-negative ratio \cite{robinson2021contrastive}. 

%Table~\ref{tab:dataset-stats} summarizes per-language statistics for four programming languages, including average Lines Of Code (LOC), total code snippets, and the average token count.

\textbf{Code retrieval.} XLCoST \cite{zhu2022xlcost} is a multilingual, parallel code benchmark designed to evaluate various code intelligence tasks in the cross-language setting. The dataset is constructed from the “Data Structures and Algorithms” section on GeeksForGeeks, where each problem is accompanied by solutions in multiple programming languages as well as the natural language description. Among the downstream tasks, code retrieval splits into two categories: NL-code retrieval and XL-code retrieval. In this paper, we focus on XL-code retrieval, i.e., given a complete program in one language as a query, retrieve functionally equivalent implementations from its parallel programs in other languages. The XLCoST dataset is divided into 100,336 training examples, 5,284 validation examples, and 9,702 test examples. To prevent data leakage from similar problems appearing in both the training and test sets, all programming problems are first clustered based on the textual similarity of their descriptions. These clusters are then split at the problem‐group level into training, validation, and test partitions in an 85\%/5\%/10\% ratio, ensuring that no two highly similar problems ever cross between sets.

\begin{comment}

\begin{table}[t]
\centering
\caption{Statistics of datasets used in experiments.}
\label{tab:dataset-stats}
\begin{tabularx}{0.65\textwidth}{@{} L r r r@{}}
    \toprule
    Language & Avg. LOC & Code snippets & Avg. Tokens \\
    \midrule
    Java   & 106 & 31,689 & 734 \\
    Python &  26 & 29,264 & 212 \\
    C\#    & 111 & 25,980 & 939 \\
    C++    &  66 & 26,918 & 616 \\
    \bottomrule
\end{tabularx}
\end{table}
\end{comment}

\subsection{Evaluation Metrics}

We regard the cross-language code clone detection task as a classification task, so we evaluate performance using Precision, Recall, and F1 score. Precision measures the proportion of true clones among all samples predicted as clones by the model. Recall measures the proportion of true clones that are correctly identified by the model. F1 score is the harmonic mean of Precision and Recall, used to provide an overall measure of performance in the classification task.

In cross-language code retrieval, we typically obtain a ranked list of relevant code snippets for each query and evaluate both the ranking quality and the hit rate. MRR and Precision@k comprehensively reflect the performance.

\subsection{Baseline Approaches}

\textbf{Code clone detection.} We compare our approach against three state-of-the-art approaches. \textbf{CLCDSA} \cite{perez2019cross} is an AST-based cross-language clone detection approach. It extracts statistical features from multilingual ASTs and leverages API documentation for the function-call information to determine the code clone relationships. \textbf{C4} \cite{tao2022c4} is a sequence-based approach built on the pre-trained CodeBERT \cite{feng2020codebert}. C4 treats code snippets in different programming languages as token sequences, fine-tunes CodeBERT using a contrastive objective, and learns to map cross-language implementations of the same programming task into a shared semantic space. \textbf{FSD-CLCD} \cite{zhang2024fsd} is a function-semantic-distillation approach for cross-language clone detection. FSD-CLCD first parses code snippets into AST graphs, applies a GMN to iteratively update both trees, uses contrastive learning to align functional semantics, and finally embeds semantically equivalent code snippets into a unified vector space.

\textbf{Code retrieval.} We compare our approach with three pre-trained Transformer baselines, including CodeBERT, UniXcoder, and GraphCodeBERT. \textbf{CodeBERT} \cite{feng2020codebert} is widely used in the XLCoST benchmark \cite{zhu2022xlcost} for cross-language code retrieval, where a dual-encoder encodes the query and candidate code separately and ranks candidates by cosine similarity after fine-tuning on positive/negative pairs. \textbf{UniXcoder} \cite{guo2022unixcoder} extends pre-training with unified code representations to better support code understanding and retrieval across different languages. \textbf{GraphCodeBERT} \cite{guographcodebert} incorporates program-structure signals (e.g., data-flow relations) to enhance the semantic modeling beyond plain token sequences. These baselines are representative and competitive for cross-language retrieval, enabling a fair comparison with our approach.

\begin{comment}
We compare our approach against three state-of-the-art approaches as follows.
\begin{enumerate}
    \item \textbf{CLCDSA} \cite{perez2019cross}: It is an AST-based cross-language clone detection approach. CLCDSA extracts statistical features from multilingual ASTs and leverages API documentation for the function-call information to determine the code clone relationships.
    \item \textbf{C4} \cite{tao2022c4}: This is a sequence-based approach built on the pre-trained CodeBERT \cite{feng2020codebert}. C4 treats code snippets in different programming languages as token sequences, fine-tunes CodeBERT using a contrastive objective, and learns to map cross-language implementations of the same programming task into a shared vector space.
    \item \textbf{FSD-CLCD} \cite{zhang2024fsd}: It is a function-semantic-distillation approach for cross-language clone detection. FSD-CLCD first parses code snippets into AST graphs, applies a GMN to iteratively update both trees, uses contrastive learning to align functional semantics, and finally embeds semantically equivalent code snippets into a unified vector space.
\end{enumerate}
\end{comment}

\subsection{Experimental Setup}

\textbf{Code clone detection.} We set the batch size to 24 for a balance between efficiency and GPU memory, with a learning rate of \(1 \times 10^{-3}\) for stable convergence. The maximum input length is 400 tokens to cover most function-level code. The model has 4 layers, with a 100-dimensional embedding. The pruning rate is set to 0.4 for moderate tree simplification, and the temperature in contrastive learning is fixed at 0.1, with a margin \(m\) of 10 to maintain distance between non-clone pairs.

\textbf{Code retrieval.} We use a multi-negative sampling combined with temperature-scaled cross-entropy strategy. For each sample, we select one positive and ten negative examples. All candidates are encoded using a GMN model with 100-dimensional node features. After encoding, we compute cosine similarities and apply a temperature coefficient of 0.1 to smooth the distribution. Training uses the Adam optimizer with an initial learning rate of \(1 \times 10^{-3}\), a batch size of 8, and runs for 10 epochs.

\subsection{Experimental Results and Analysis}

\noindent\textbf{RQ1: Can our approach achieve better results than the baseline approaches?}

\textbf{Code clone detection.} We conduct the comparison experiment based on two scenarios. In the first scenario, we consider all the programming languages in the benchmark dataset and evaluate the performance of all the approaches in the multilingual scenario. In contrast, in the second scenario, we consider one to one (and vice versa) cross-language clone detection, e.g., Java to Python and Python to Java code clone detection.

The multilingual code clone detection results are shown in Table \ref{tab:clone-results-1}. As a whole, our approach outperforms all existing baselines in terms of Precision, Recall, and F1 score. For example, CLCDSA achieves 50.35\%, 91.92\%, 65.60\% for Precision, Recall, F1 score respectively. C4 achieves better results than CLCOSA. C4 achieves 94.34\%, 90.21\%, 92.16\% in terms of the three evaluation metrics. Even though FSD-CLCD already achieves a good performance, i.e., 96.52\%, 97.72\%, 96.94\%, our approach further boosts these metrics to 99.94\%, 99.92\%, 99.93\% for the three evaluation metrics. It means that our approach not only eliminates syntactic discrepancies across programming languages but also precisely captures functionally equivalent semantic implementations, thus effectively minimizing both false positives and false negatives.

\begin{table}[t]
  \centering
  \caption{The multilingual code clone detection results}
  \label{tab:clone-results-1}
  \begin{tabularx}{0.45\textwidth}{@{}lYYY@{}}
    \toprule
    Approach      & Precision & Recall & F1 Score    \\
    \midrule
    CLCDSA      & 50.35\%      & 91.92\%   & 65.62\%   \\
    C4          & 94.34\%      & 90.21\%   & 92.16\%   \\
    FSD‐CLCD    & 95.62\%      & 97.72\%   & 96.94\%   \\
    Our approach        & \textbf{99.94\%} & \textbf{99.92\%} & \textbf{99.93\%} \\
    \bottomrule
  \end{tabularx}
\end{table}

Furthermore, when we evaluated our approach in all the one to one language pairs (Java <--> Python, Java <--> C++, Java <--> C \#, Python <--> C++, Python <--> C \#, and C++ <--> C \#), our approach maintains exceptional generalization, as shown in Table \ref{tab:clone-results}. In all the one to one language pairs, our approach achieves Precision $\geq$ 99.70\%, Recall $\geq$ 99.83\%, and F1 score $\geq$ 99.81\%. In static-typed or mixed static-dynamic language pairs (e.g., Java <--> Python, Java <--> C++, Java <--> C \#), F1 score achieved by our approach reaches to at least 99.93\%. Even for the most syntactically divergent language pair, i.e., Python <--> C++, F1 score achieved by our approach remains a robust 99.81\%. This demonstrates that, regardless of the language pairs, the synergy of unified AST abstraction and cross‐graph attention in our approach consistently pulls functionally equivalent code snippets into the same vector space, showing strong adaptability. 

To rule out data leakage and overfitting, we conducted rigorous checks. On the data side, we performed a split-integrity checking, confirming no overlap in Task IDs, code instance identifiers (ID1/ID2), or unordered code pairs between training and test splits. On the model side, we compared validation and test performance and found no evidence of overfitting. Detailed verification results are provided in the (\emph{Discussion}) Section.

\begin{table}[t]
  \centering
  \caption{One to one cross‐language clone detection results}
  \label{tab:clone-results}
  \begin{tabularx}{0.45\textwidth}{@{}lYYY@{}}
    \toprule
    One to One Language Pair      & Precision & Recall & F1 Score     \\
    \midrule
    Java <--> Python       & 100.00\%      & 99.88\%   & 99.94\%   \\
    Java <--> C++          & 99.93\%      & 100.00\%   & 99.96\%   \\
    Java <--> C\#          & 99.87\%      & 100.00\%   & 99.93\%   \\
    Python <--> C++        & 99.70\%      & 99.92\%   & 99.81\%   \\
    Python <--> C\#        & 100.00\%      & 99.87\%   & 99.93\%   \\
    C++ <--> C\#           & 100.00\%      & 99.83\%   & 99.91\%   \\
    \bottomrule
  \end{tabularx}
\end{table}

\textbf{Code retrieval.} As shown in Table~\ref{tab:retrieval-results}, compared with CodeBERT \cite{feng2020codebert}, UniXcoder, and GraphCodeBERT on the same dataset, our approach achieves significant improvements in both MRR and Precision@4. In this table, "Language" refers to the code snippets of this specific programming language being used as the query, with the remaining programming languages used for retrieval. Since the benchmark dataset involves five programming languages in total, when using one language as the query, there are at most four semantically equivalent implementations available in the remaining programming languages. Therefore, Precision@4 is adopted, as it exactly corresponds to the maximum number of correct answers that can be retrieved for each query. Among the baselines, UniXCoder achieves the best average MRR and Precision@4, so we primarily compare our approach against UniXCoder. Specifically, MRR achieved by UniXCoder ranges from \textbf{0.4380} to \textbf{0.5131}, with an overall average of \textbf{0.4909}. In contrast, our approach raises the average MRR to \textbf{0.5547}, yielding an absolute gain of \textbf{0.0638} (approximately \textbf{13.0\%}). The improvements are consistent across all query languages, with the largest gain observed when using \textbf{C++} as the query (\textbf{+0.0839}), followed by \textbf{Java} (\textbf{+0.0678}) and \textbf{C\#} (\textbf{+0.0664}); \textbf{Python} and \textbf{C} queries also improve by \textbf{+0.0549} and \textbf{+0.0458}, respectively. This consistent increase indicates stable cross-language generalization of the shared semantic space and a stronger capability to rank multiple correct answers near the top of the retrieval list.

In terms of \textbf{Precision@4}, UniXCoder also provides the best performance in the baselines, reaching an average of \textbf{91.26\%}. Our approach further improves this metric to \textbf{93.78\%} on average (\textbf{+2.52} percentage points), and outperforms UniXCoder for every query language: \textbf{C} (91.34\% vs.\ 83.39\%), \textbf{Python} (91.62\% vs.\ 90.93\%), \textbf{C++} (94.85\% vs.\ 93.12\%), \textbf{Java} (96.82\% vs.\ 95.23\%), and \textbf{C\#} (94.29\% vs.\ 93.61\%). Notably, our approach achieves nearly \textbf{97\%} Precision@4 when querying in Java. Although CodeBERT slightly exceeds our Precision@4 when \textbf{C} is used as the query (93.73\% vs.\ 91.34\%), its performance drops substantially on the other languages, whereas our approach maintains consistently high Precision@4 across all programming languages.

\begin{table}[t]
  \centering
  \caption{Comparison of MRR and Precision@4 between our approach and baselines.}
  \label{tab:retrieval-results}

  \setlength{\tabcolsep}{4pt}
  \resizebox{0.95\linewidth}{!}{%
  \begin{tabular}{llcccc}
    \toprule
    Language & Metric & Our approach & CodeBERT & UniXcoder & GraphCodeBERT \\
    \midrule
    \multirow{2}{*}{C} % 修正了这里
      & MRR & \textbf{0.4838} & 0.4117 & 0.4380 & 0.4371 \\
      & Precision@4 (\%) & 91.34\% & \textbf{93.73\%} & 83.39\% & 84.29\% \\
    \midrule
    \multirow{2}{*}{Python} % 修正了这里
      & MRR & \textbf{0.5572} & 0.4831 & 0.5023 & 0.2041 \\
      & Precision@4 (\%) & \textbf{91.62\%} & 74.25\% & 90.93\% & 38.69\% \\
    \midrule
    \multirow{2}{*}{C++} % 修正了这里
      & MRR & \textbf{0.5725} & 0.4882 & 0.4886 & 0.4376 \\
      & Precision@4 (\%) & \textbf{94.85\%} & 73.72\% & 93.12\% & 83.20\% \\
    \midrule
    \multirow{2}{*}{Java} % 修正了这里
      & MRR & \textbf{0.5804} & 0.4874 & 0.5126 & 0.5165 \\
      & Precision@4 (\%) & \textbf{96.82\%} & 73.70\% & 95.23\% & 95.89\% \\
    \midrule
    \multirow{2}{*}{C\#} % 修正了这里
      & MRR & \textbf{0.5795} & 0.4876 & 0.5131 & 0.5195 \\
      & Precision@4 (\%) & \textbf{94.29\%} & 74.33\% & 93.61\% & 93.53\% \\
    \midrule
    \textbf{Average}
      & MRR & \textbf{0.5547} & 0.4716 & 0.4909 & 0.4230 \\
      & Precision@4 (\%) & \textbf{93.78\%} & 77.95\% & 91.26\% & 79.12\% \\
    \bottomrule
  \end{tabular}%
  }
\end{table}

\noindent\textbf{RQ2: Is the hard-negative instance construction strategy effective in our approach?}

\begin{figure}[t]
    \centering
    \includegraphics[width=\linewidth]{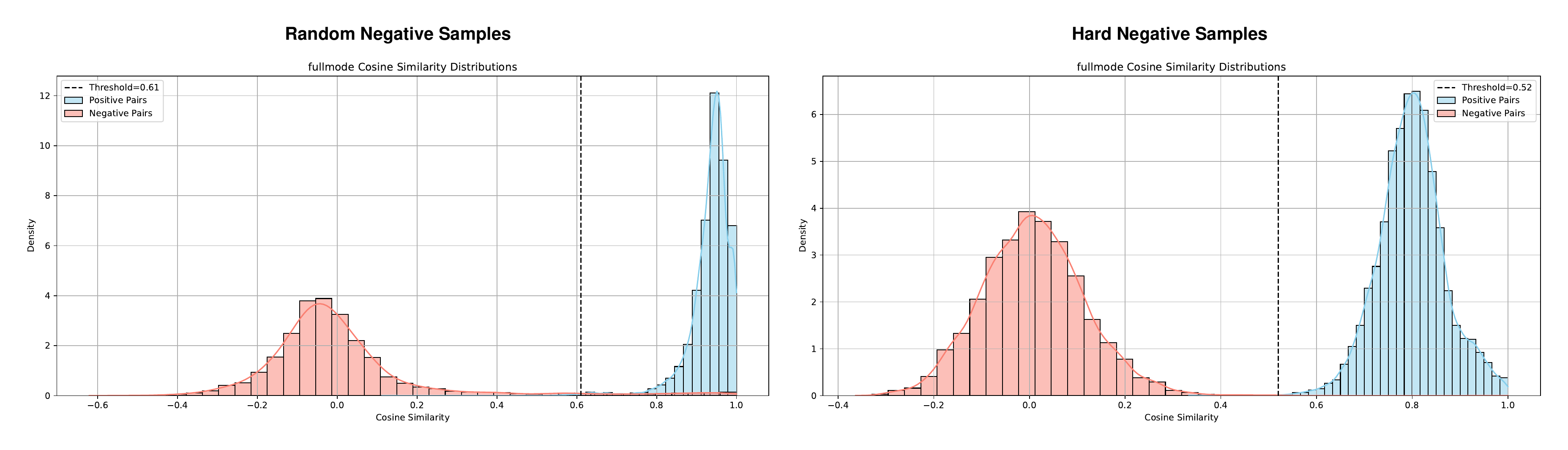}
    \caption{The difference between hard-negative mining and random negative sampling}
    \label{fig:hard-negative}
\end{figure}

\textbf{Code clone detection.} In this RQ, we evaluate the negative instance construction strategy in our approach by comparing it against a random construction strategy, which randomly selects a code snippet from another programming language to form a cross-language code clone negative pair. We conduct the experiment in the multilingual scenario. As shown in the Figure~\ref{fig:hard-negative}, we display the cosine similarity distributions for positive cross-language code clone pairs (blue) and negative cross-language code cline pairs (red), along with the selected decision threshold (black dashed line). The gap between these distributions illustrates how the two negative instance construction strategies influence the training of our approach.

%Considering the random negative instance construction strategy, the randomly selected code snippets from another programming language bear almost no relation to each other in the shared semantic space. Their cosine similarities cluster around zero, with some even falling into negative values. Positives instance, on the other hand, almost all lie in the high-similarity range from 0.8–1.0, creating a clear separation in the histogram with virtually no overlap. In this case, setting a low threshold (e.g., 0.3) easily yields nearly 100\% Precision and Recall on the training set, which indicates that positives and negatives under random sampling are almost trivially separable. However, this phenomenon may overestimate the model’s capability, as it does not necessarily demonstrate that the model has captured deep cross-language semantic alignment.

When we adopt the hard-negative construction strategy, the negative distribution shifts closer to the positives. Code clone pairs that are structurally and label-wise very similar in the unified AST space but are non-code-clones end up with relatively high cosine similarities after encoding. The right tail of the red (negative) distribution moves into the range of 0.2–0.6 or higher, overlapping substantially with the blue (positive) distribution. Only when the model actually learns to distinguish these superficially indistinguishable but semantically different samples can it maintain both high Precision and high Recall at a higher decision threshold.

\begin{table}[t]
\centering
\caption{MRR under Different \texttt{neg\_k} Values}
\label{tab:neg_K}
\begin{tabular}{cccccc}
\hline
 & C & Python & C++ & Java & C\# \\
\hline
\texttt{neg\_k=0} & \textbf{0.6468} & 0.0256 & 0.0468 & 0.2123 & 0.1946 \\
\texttt{neg\_k=1} & 0.4655 & 0.5458 & 0.5558 & 0.5729 & 0.5666 \\
\texttt{neg\_k=3} & 0.4827 & 0.5423 & 0.5649 & 0.5795 & 0.5685 \\
\texttt{neg\_k=5} & 0.4792 & 0.5526 & 0.5698 & 0.5801 & 0.5731 \\
\texttt{neg\_k=10} & 0.4838 & \textbf{0.5572} & \textbf{0.5725} & \textbf{0.5804} & \textbf{0.5795} \\
\hline
\end{tabular}
\end{table}

\textbf{Code retrieval.} In terms of code retrieval, we examined how the number of negative samples (neg\_k) affect the performance of our approach in cross‐language code retrieval. Table~\ref{tab:neg_K} show the impacts of different neg\_k values on the result of MRR achieved by our approach. Here, neg\_k = 0 means that no negative samples are used during training (i.e., the model is only exposed to positive pairs), while neg\_k = 10 indicates that each positive pair is accompanied by ten negative samples for contrastive learning. Similarly, a specific programming language in this table means that we use the code snippets in this programming language as the queries to retrieve the code snippets from other programming languages. From the results, we found that with no negative samples (neg\_k=0), MRR is extremely low across all programming languages, indicating our approach lacks a clear decision boundary. Introducing just one negative sample (neg\_k=1) causes performance to increase dramatically, and MRR continues to improve as neg\_k increases, peaking at neg\_k=10 with an average gain of about 0.02. Importantly, the gains start to plateau between k=5 and k=10, suggesting that our approach has already received sufficient negative contrast information in this range.

\noindent\textbf{RQ3: What is the effect of each module on the performance of our approach?}

\begin{table}[t]
  \centering
  \caption{Ablation study results on cross-language code clone detection}
  \label{tab:ablation-clone}
  \begin{tabular}{lccc}
    \toprule
    Setting                           & Precision & Recall  & F1 Score     \\
    \midrule
    w/o hard‐negative construction        & 55.24\%    & 99.59\%  & 71.06\%  \\
    w/o cross‐graph attention         & 71.89\%    & 81.62\%  & 76.45\%  \\
    w/o AST enhancement               & 94.53\%    & 98.67\%  & 96.55\%  \\
    Full model                        & \textbf{99.94\%} & \textbf{99.92\%} & \textbf{99.93\%} \\
    \bottomrule
  \end{tabular}
\end{table}

\textbf{Code clone detection.} In this experiment, we remove the hard-negative construction strategy, the cross-graph attention, and the AST enhancement modules individually to show their impacts. When we remove a specific module, the other modules are remained the same. In such a way, we can measure the influence of the module on the final performance of our approach. The results are shown in Table~\ref{tab:ablation-clone}.

From the results, removing hard-negative construction causes performance to collapse: while Recall stays high (99.59\%), Precision drops to 55.24\%, yielding an F1 of only 71.06\%. This indicates that training with positives alone provides no signal for what is \emph{not} a clone, so the model predicts most pairs as clones and the decision boundary effectively disappears. Removing cross-graph attention further reduces Precision/Recall to 71.89\%/81.62\% (F1 76.45\%). In this case, the model can only aggregate information within each AST and fails to capture cross-language correspondences, leading to missed true clones and poor rejection of local “pseudo-clones”. Finally, removing AST structural enhancement improves Precision/Recall to 94.53\%/98.67\% (F1 96.55\%), but still lags behind the full model. Structural enhancement reduces noise and guides cross-graph attention toward key control-flow and declaration nodes; without it, redundant and noisy nodes distract the model and weaken discrimination.

The performance of our approach gains arise from the joint effect of the three complementary components, i.e., the unified AST abstraction that preserves key semantic structures while reducing noise, the structural enhancement that stabilizes cross-graph alignment, and the GMN that explicitly models semantic interactions across programming languages. Combined with a hard-negative construction strategy to sharpen discriminative power, these modules collectively improve Precision, Recall, and F1 score, yielding a reliable and generalizable solution for cross-language code clone detection.

\textbf{Code retrieval.} In this experiment, we evaluate the impact of two core modules, i.e., the AST enhancement and the cross‐graph attention on the performance of our approach in cross‐language code retrieval. Similarly, we remove the two modules individually and keep the other modules the same to obtain their impacts. The MRR and Precision@4 results are listed in Table~\ref{tab:ablation-retrieval}. The specific programming language in this table means that we use the code snippets in this programming language as the queries to retrieve the code snippets from other programming languages.

From the results, we can see that removing the AST enhancement causes MRR achieved by our approach to fall by about 0.015 on average across all programming languages, with Precision@4 also dipping by roughly 5\%. It means that lacking a unified global‐root anchor and noise‐node pruning, our approach struggles to pinpoint key functional substructures in the high‐noise raw AST, which degrades both MRR and Precision. Similarly, removing the cross‐graph attention leads to an average MRR decrease of around 0.012 and a Precision@4 drop of approximately 3\%–4\% of our approach. Without node‐level interaction with candidate graphs, our approach loses its ability to capture deep semantic correspondences in different programming languages, making it hard to tightly cluster queries with functionally equivalent candidates and resulting in lower MRR and Precision.

These gains reflect the strength of our unified AST combined with GMN cross-graph alignment strategy. Unified AST abstraction removes interlanguage node discrepancies, while cross-graph attention and multi-round GRU updates enable explicit, node-level information exchange to capture deep functional equivalence. In the multi-answer retrieval scenario, this fine-grained alignment lets multiple correct results from different programming languages all rank near the top. In contrast, CodeBERT lacks specialized modeling of cross-language AST structure and therefore underperforms in multilingual, multi-answer retrieval. The results demonstrate that integrating the cross-language syntax information with neural alignment is essential for accurate and comprehensive code retrieval.

\begin{table}[t]
  \centering
  \caption{Ablation study results for cross‐language code retrieval}
  \label{tab:ablation-retrieval}
  \resizebox{\linewidth}{!}{%
  \begin{tabular}{lccccc}
    \toprule
    & \multicolumn{5}{c}{MRR} \\
    \cmidrule(lr){2-6}
    Configuration         & C      & Python & C++    & Java   & C\#    \\
    \midrule
    w/o AST enhancement       & 0.4778 & 0.5364 & 0.5558 & 0.5685 & 0.5688 \\
    w/o Cross‐graph attention & 0.4699 & 0.5565 & 0.5692 & 0.5768 & 0.5706 \\
    Full model                & \textbf{0.4838} & \textbf{0.5572} & \textbf{0.5725} & \textbf{0.5804} & \textbf{0.5795} \\
    \midrule
    & \multicolumn{5}{c}{Precision@4} \\
    \cmidrule(lr){2-6}
    Configuration         & C      & Python & C++    & Java   & C\#    \\
    \midrule
    w/o AST enhancement       & 89.05\% & 84.56\% & 91.92\% & 96.13\% & \textbf{94.92\%} \\
    w/o Cross‐graph attention & 89.40\% & 91.10\% & 94.36\% & 96.35\% & 93.85\% \\
    Full model                & \textbf{91.34\%} & \textbf{91.62\%} & \textbf{94.85\%} & \textbf{96.82\%} & 94.29\% \\
    \bottomrule
  \end{tabular}%
  }
\end{table}

\section{Discussion}

\subsection{Semantic Space Visualization of Our Approach}

\begin{figure}[t]
    \centering
    \begin{subfigure}[b]{0.49\textwidth}
        \includegraphics[width=\textwidth]{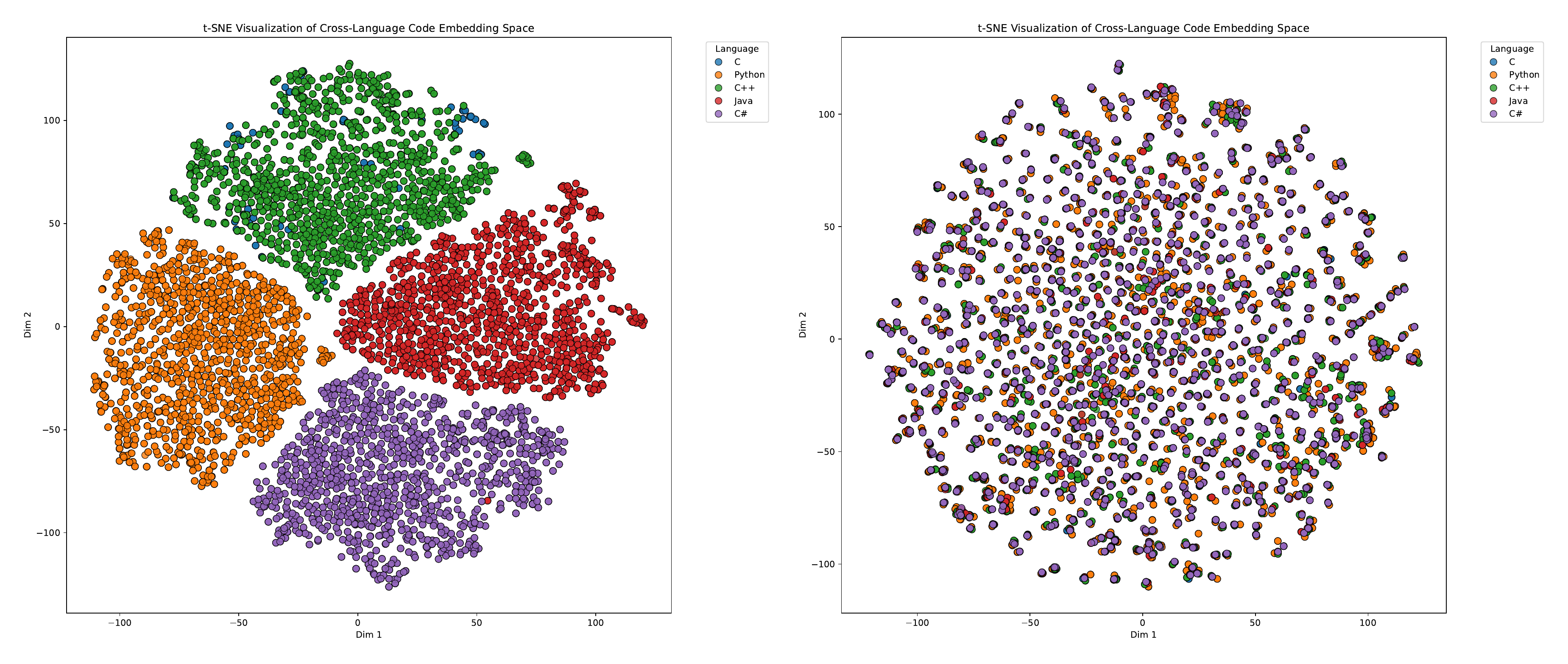}
        \caption{t-SNE visualization based on programming language level embeddings, showing the distribution of multiple programming languages before (left) and after (right) training.}
        \label{fig:tsne-1}
    \end{subfigure}
    \hfill % 
    \begin{subfigure}[b]{0.49\textwidth}
        \includegraphics[width=\textwidth]{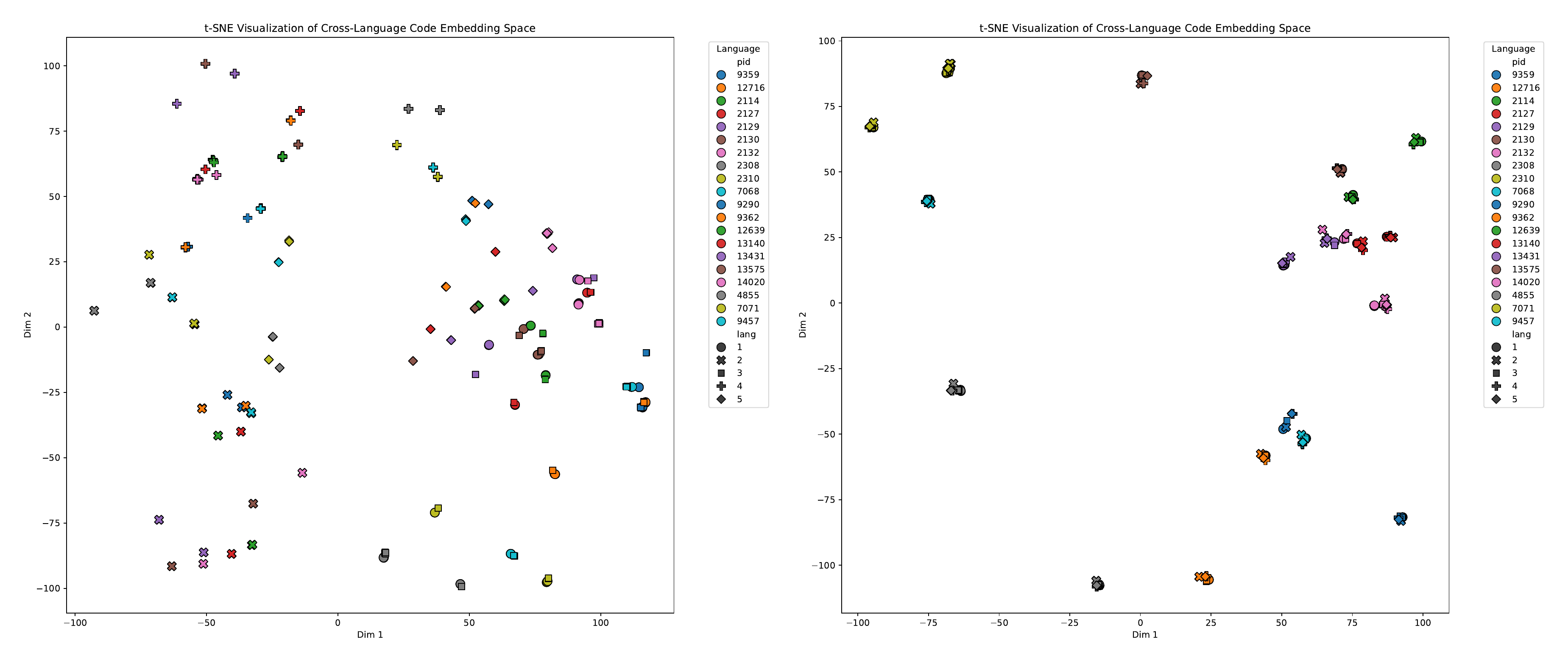}
        \caption{t-SNE visualization of code snippets from different languages implementing the same task, which are scattered before training but form tight clusters after training.}
        \label{fig:tsne-2}
    \end{subfigure}
    \caption{Visualization of cross-language code embeddings before and after training}
    \label{fig:tsne}
\end{figure}

To evaluate our shared semantic space in terms of language agnosticism and semantic cohesion, we performed a visualization of the embeddings of multilingual code snippets. Specifically, we used all 9,702 test examples from the XLCoST dataset \cite{zhu2022xlcost}, extracted their vector representations using our approach, and applied t-SNE \cite{van2008visualizing} to project these high-dimensional embeddings into two dimensions. In Figure~\ref{fig:tsne-1}, colors denote the programming language of each code snippet. As illustrated in the comparison, the left plot shows the 2D t-SNE projection of the same embeddings of cross-language code snippets before the training. We can see that each programming language forms its own distinct cluster, indicating that with only the initial mapping of the unified label, the vectors still heavily depend on syntactic language differences and remain clearly separated. In contrast, the right plot shows the projection after the full training of our approach. We can see that the points of all the programming languages are now highly intermixed, with no visible language boundaries. This transformation directly demonstrates the design goal of our semantic space, i.e., cross-language functional equivalence has been effectively aligned. Our approach groups multilingual implementations by actual functionality in a common subspace.

Figure~\ref{fig:tsne-2} further validates the semantic cohesion of the shared space by marking points with the task ID and the programming language. In the left plot, projected after only unified AST abstraction and before any GMN training, points from the same task but different programming languages are widely scattered, and overall each language still forms fragmented clusters, indicating that the embeddings at this stage remain heavily influenced by syntax and language-specific features rather than the functionality. In contrast, the right plot, after end-to-end training, shows that all multilingual implementations of the same task collapse into tight clusters. Code snippets no longer group by the programming languages but instead intermingle by the tasks, with minimal intracluster distances. Globally, distinct clusters are well separated in the 2D plane, yet within each cluster points are highly consistent, demonstrating that the shared space of our approach successfully brings together cross-language code for the same functionality while keeping different tasks apart.

This visualization demonstrates that, after training with the unified AST abstraction and the GMN-based cross‐graph attention, our approach achieves deep functional alignment across programming languages. Implementations of the same task are mapped to adjacent regions regardless of the programming languages, while different tasks remain in distinct clusters, preserving semantic separation. These semantic clusters provide a robust foundation for the understanding and manipulation of cross-language code snippets.

\subsection{Threats To Validity}

\textbf{Internal validity.} One potential threat lies in how the training and test data are constructed. Misalignment or annotation noise may still exist and could affect evaluation reliability. Given the near-ceiling results in Table~\ref{tab:clone-results}, we further checked whether performance might be inflated by data leakage or overfitting. We conducted a strict split-integrity checking and confirmed \emph{zero overlap} between training and test splits in (i) Task IDs and (ii) the union of code instance IDs (ID1/ID2). The checking verified that the training dataset contains 69,434 pairs over 1,093 tasks and the test dataset contains 8,699 pairs over 137 tasks, ruling out trivial leakage from duplicated tasks, code instances, or repeated pairs across splits. While this does not fully exclude near-duplicate leakage or biases from empirical hyperparameter choices, it substantially reduces this concern. To assess overfitting, we compared validation and test behavior and found them highly consistent. The F1 score is 0.9995 on validation and 0.9993 on test, indicating stable generalization rather than train-only memorization.

\textbf{External validity.} Our experiments were conducted on the XLCoST dataset, which covers five programming languages. Although it is the largest available multilingual code dataset, it does not fully represent all programming paradigms or real-world industrial codebases. The generalizability of our approach to low-resource languages, domain-specific languages, or large-scale software systems remains to be further validated. Future work should extend the evaluation to a broader range of datasets and programming languages to further assess the robustness and applicability of our approach.

\section{Related Work}

\textbf{Construction of the Shared Semantic Vector Space.} Cross-language code understanding increasingly focuses on mapping code to a shared semantic space \cite{han2021learning}. While early word embedding alignment methods \cite{mikolov2013efficient, lample2018word} laid the groundwork, they operate at the text level and ignore code structure. Recent GNN-based approaches \cite{zhang2025ast, rahaman2024source} address this by converting ASTs into graphs, yet they typically encode languages separately, lacking an interactive cross-graph alignment mechanism \cite{wang2022unified}. Our approach unifies AST node labels and uses a GMN with cross-graph attention to embed code into a shared space. This captures functional equivalence more effectively than sequence or single-graph baselines by preserving structural information.

\textbf{Cross-Language Code Clone Detection.} Previous approaches \cite{cheng2017clcminer,perez2019cross} largely relied on statistical features or API documentation. While recent works like C4 \cite{tao2022c4} and FSD-CLCD \cite{zhang2024fsd} incorporate deep learning, they primarily treat code as sequences or linearized ASTs. Conversely, our study unifies AST labels and structures, utilizing a GMN to explicitly model node-level semantic alignment. This graph-based approach significantly outperforms sequence-based methods in Precision and Recall.

\textbf{Cross-Language Code Retrieval.} While CodeBERT \cite{feng2020codebert, guographcodebert} excels in monolingual settings, cross-language retrieval is less developed \cite{mathew2021cross}. Current methods fine-tuned on XLCoST \cite{zhu2022xlcost} typically neglect AST structural diversity. In contrast, our approach explicitly maps multilingual code to a unified semantic space and employs cross-graph alignment, outperforming baselines in MRR and Precision@K.

\section{Conclusion}

In this paper, we propose an approach that integrates a unified AST abstraction and GMN‐based cross‐graph alignment to bridge the gap in understanding multilingual code snippets. By mapping the ASTs of code snippets from different programming languages to a set of common labels with structural enhancement and applying cross‐graph attention, our approach embeds functionally equivalent ASTs in a shared semantic space. We apply our approach in two cross-language tasks, i.e., code clone detection and code retrieval. Experimental results show that our approach achieves better results than the existing state-of-the-art approaches.

\section*{Acknowledgment}
This work was supported by the National Natural Science Foundation of China under Grant No. 62272225 and the Open Fund of the Shanghai Key Laboratory of Space-based Heterogeneous Network Collaborative Computing.

%\section*{Data Availability}

%The datasets used in this study are publicly available (Google CodeJam \cite{codejam}, 
%AtCoder \cite{atcoder}, and XLCoST~\cite{zhu2022xlcost}). The replication package, including data preprocessing scripts, trained models, and evaluation code, is openly available at \url{https://github.com/CenOspreay/BPLangGap}. This ensures transparency and facilitates reproducibility of the experimental results.

\bibliographystyle{ACM-Reference-Format}
\bibliography{references}

\end{document}